\documentclass[a4paper,12pt]{article}
\usepackage{amsmath,amsfonts,amssymb,epsfig,subfigure,color}

\usepackage{caption}
\usepackage[percent]{overpic}

\renewcommand{\ge}{\geqslant}

\newcommand{\be}{\begin{equation}}
\newcommand{\en}{\end{equation}}

\begin{document}

\numberwithin{equation}{section}


\title{Gent models for the inflation\\ of spherical balloons\footnote{
\emph{In memory of Alan Gent, with gratitude for all his inspiring papers}}}


\author{Robert Mangan, Michel Destrade\\[12pt]
  School of Mathematics, Statistics and Applied Mathematics,  \\ 
   National University of Ireland Galway, \\University Road, Galway, Ireland}

\date{}

\maketitle

\begin{abstract}

We revisit an iconic deformation of nonlinear elasticity: the inflation of a rubber spherical thin shell.
We use the 3-parameter Mooney and  Gent-Gent (GG) phenomenological models to explain the stretch-strain curve of a typical inflation, as these two models cover a wide spectrum of known models for rubber, including the Varga, Mooney-Rivlin, one-term Ogden, Gent-Thomas and Gent models.
We find that the basic physics of inflation exclude the Varga, one-term Ogden and Gent-Thomas models.
We find the  link between the exact solution of nonlinear elasticity and the membrane and  Young-Laplace theories often used \emph{a priori} in the literature. 
We compare the performance of both models on fitting the data for experiments on rubber balloons and animal bladder.  We conclude that the GG model is the most accurate and versatile model on offer for the modelling of rubber \color{black} balloon inflation. \color{black}

\end{abstract}

\noindent
\textbf{keywords:}
 spherical shells; membrane theory; rubber modelling; Gent model


\section{Introduction}

\begin{figure}[htp] %
\centering
\subfigure[Inflation of  a rubber balloon]{%
\includegraphics[width=0.5\textwidth]{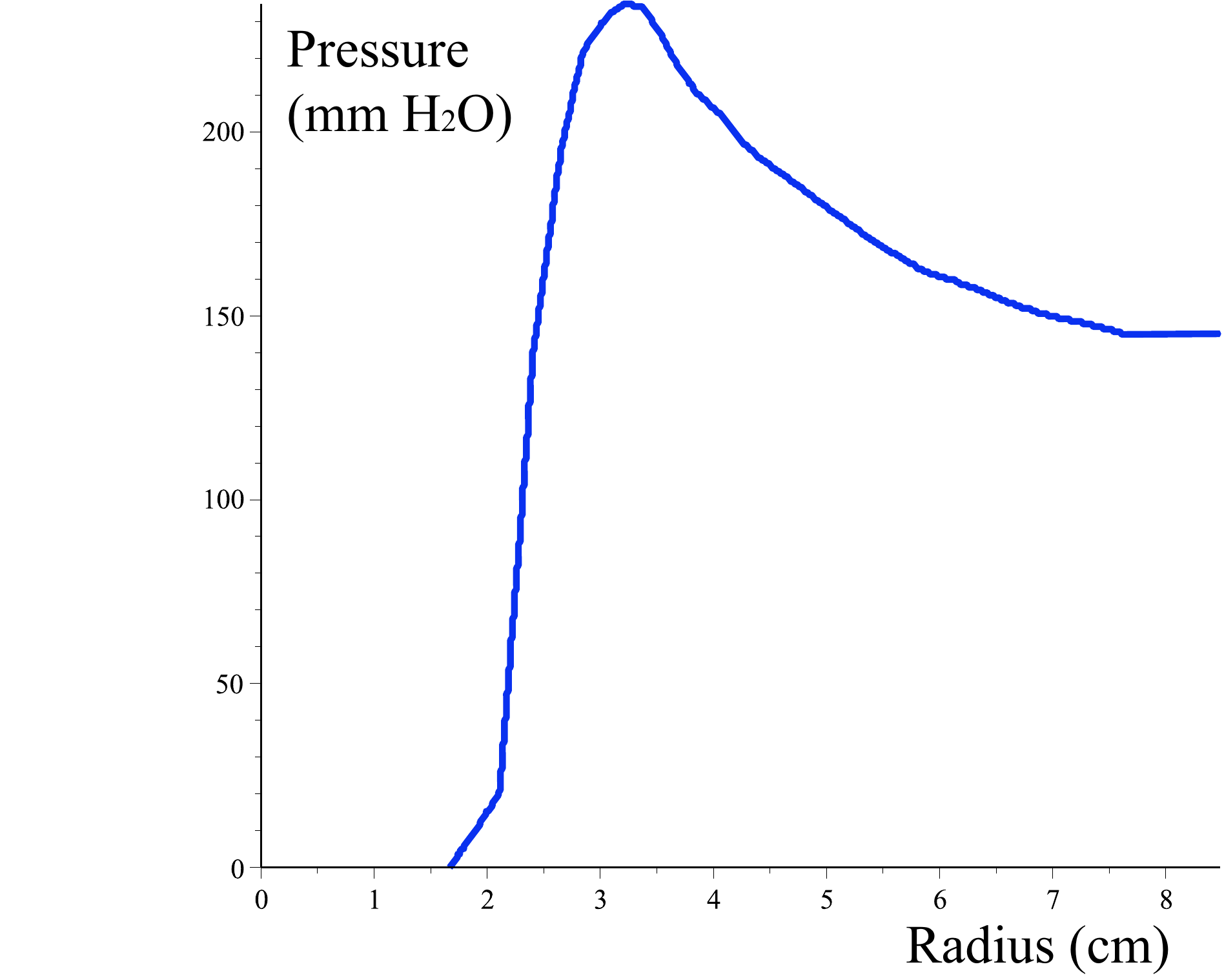}%
\label{osborne-rubber}%
}\hfil
\subfigure[Inflation of a monkey bladder]{%
\includegraphics[width=0.5\textwidth]{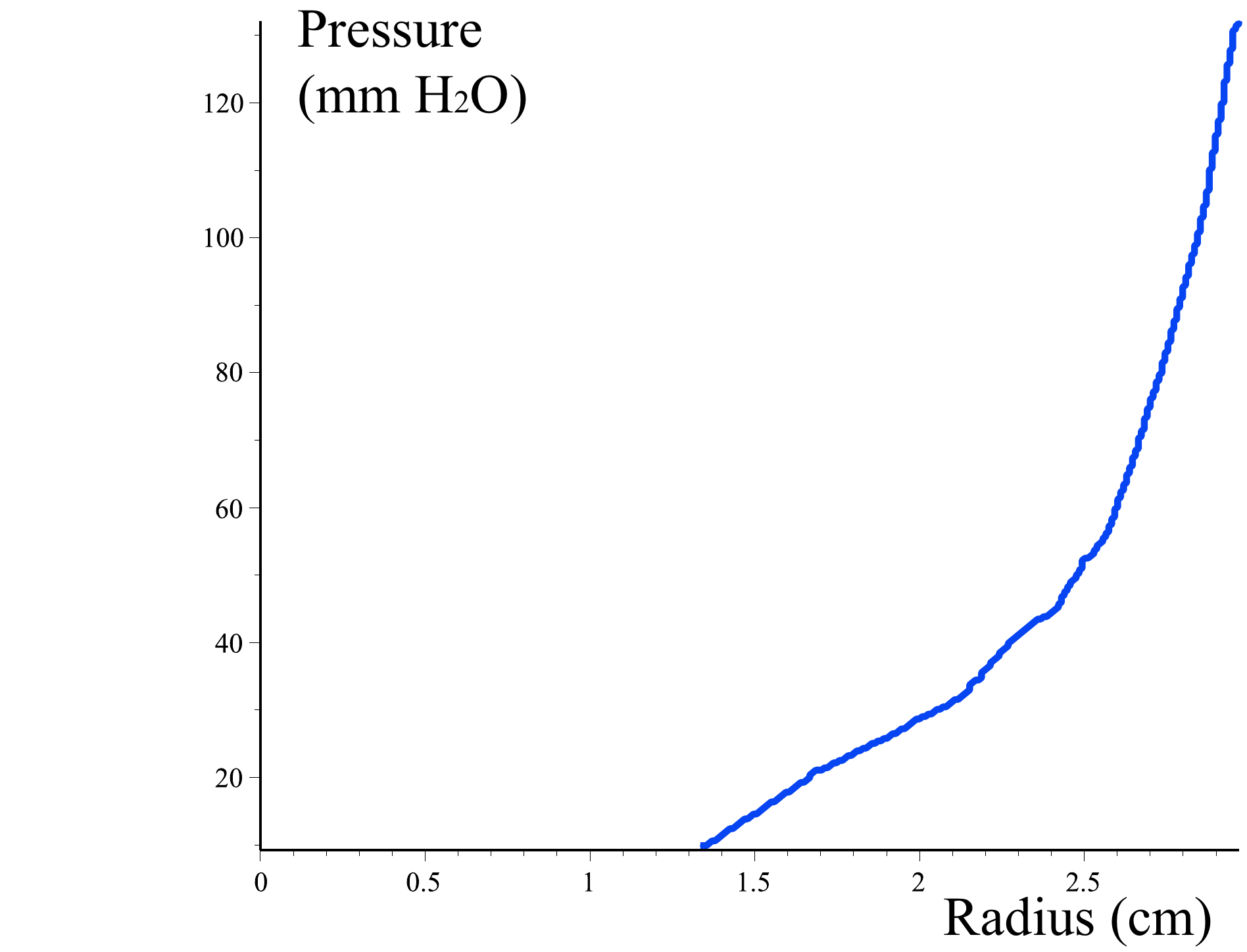}%
\label{bladder}%
}
\caption{Pressure-radius curves, digitised from Osborne's 1909 article \cite{osborne} \label{osborne}}
\end{figure}

An early success of the modern theory of nonlinear elasticity, as initiated by Ronald Rivlin in the 1950s, has been the satisfactory modelling of the inflation of a spherical shell. As early as 1909, Osborne \cite{osborne} noticed that rubber balloons and monkey bladders had a completely different mechanical response to inflation. As is known to anyone who has blown up a rubber party balloon, the initial inflation requires a strong effort, followed by an easing of the pressure required to continue until a new stiffening regime is entered, going all the way to rupture. 
In contrast, the pressure-radius graph for balloons made of biological tissue (such as early footballs) has a J shape  corresponding to an ever increasing, monotonic, effort. 
Figure \ref{osborne} reproduces Osborne's classic results (we digitised the graphs displayed in the original 1909 article).

For incompressible isotropic materials, inflation is a universal solution and thus any strain energy density can be a candidate to model the behaviour of a real blown-up material. 
In this paper we revisit some salient features of this inhomogeneous solution. 
Starting from the exact solution of non-linear elasticity for inflated finite thickness spherical shells, we recover and justify rigorously some of the assumptions made sometimes \emph{a priori} for thin elastic membranes. 
For instance in linear elasticity it is often assumed that the normal stress component in the membrane is small compared to the circumferential stress components: here we show that when expanded, the ratio of the former to the latter is of order one in the relative shell thickness $\delta = B/A -1$, where $A$, $B$ are the inner and outer initial radii, respectively. 
Similarly, for modelling rubber balloons, it is often assumed that the Young-Laplace equations for the equilibrium of bubbles should apply. 
Here we provide a rigorous basis for making that assumption.
These connections are derived in Section \ref{derivation}, and then repeated in the Appendix for the case of cylindrical shells.

Using \color{black}the approximation for the radial stress, \color{black}we then look for strain energy densities which would give a reasonable fit of some classic  data to phenomenological models.
Hence in Section \ref{curves} we evaluate the performance of two 3-parameter models, which more or less cover the entire known spectrum of stress-strain response for rubber-like materials, including the neo-Hookean,  Mooney-Rivlin,  Varga, one-term Ogden, Gent-Thomas and Gent models.
\color{black}We establish explicitly and/or numerically the limitations to be imposed on the material parameters to predict reasonable physical behaviour.\color{black}


\section{Derivation of the pressure-stretch relationship}
\label{derivation}


Here we recall the exact equations governing the equilibrium of spherical shells subject to hydrostatic pressure (see e.g., Ogden \cite{Ogden}) and then specialise them to the case of a thin membrane. 

Consider a spherical shell made of an incompressible isotropic hyperelastic material. 
\color{black}Let $\mathbf{X}= \mathbf{X}(R, \Theta, \Phi)$ and $\mathbf{x}= \mathbf{x}(r, \theta, \phi)$ \color{black} denote the  position of a material particle in the reference and current configurations, respectively. 
The associated orthonormal bases are $(\mathbf{E}_1,\mathbf{E}_{2},\mathbf{E}_{3})$ and $(\mathbf{e}_1,\mathbf{e}_{2},\mathbf{e}_{3})$, respectively.  
Suppose the shell is subject to a uniform internal pressure $P$ and assume it retains its spherical symmetry as it inflates.  
Then, \color{black} for purely radial deformations, \color{black} the motion of a particle in the shell can be described by
\begin{equation}\label{motion}  r = r(R), \quad \theta = \Theta, \quad \phi = \Phi  .\end{equation}
From (\ref{motion}), we find that the   deformation gradient, $\mathbf{F} = \partial \mathbf{x}/\partial \mathbf{X}$,  is given by
\begin{equation}\mathbf{ F} = \mathrm{diag}(dr/dR, r/R, r/R)  , \end{equation}
in the ($\mathbf{e}_i \otimes\mathbf{E}_j$) basis.
The stretch is the ratio of the length of a line element in the current configuration to the length of the corresponding line element in the reference configuration. The principal stretches are the  square roots of the eigenvalues of the left Cauchy-Green tensor, $\mathbf{B = FF^T}$, and the principal directions are along the corresponding eigenvectors. Here, the principal stretches are thus
\begin{equation}   
{\lambda}_1 = \frac{\text dr}{\text dR}  \qquad \mathrm{and} \qquad{\lambda}_{2} = {\lambda}_{3} = \frac{r}{R}  , \end{equation}
and the corresponding principal directions are along  $\mathbf{e}_1,\mathbf{e}_{2}$ and $\mathbf{e}_{3}$, respectively. Here $\lambda_1$  is the \textit{radial stretch} and $\lambda_2$ is the \textit{circumferential stretch}. 
Using the incompressibility condition, $\det{\mathbf{ F}} = {\lambda}_{1} {\lambda}_{2}{\lambda}_3=1 $,  we have 
\begin{equation} \label{lambdar}
\frac{\text dr}{\text dR}= \frac{R^2}{r^2}. \end{equation}

Letting $A$, $B$ and $a$, $b$ denote the inner radius and outer radius of the shell in the reference and current configurations, respectively,  and solving (\ref{lambdar}) leads to 
\begin{equation}
R^3 - A^3 = r^3 - a^3 , \quad R^3 - B^3 = r^3 -b^3 , \end{equation}
so that
\begin{equation} \label{lambdaa}
1 - {{\lambda}_a}^3 = \frac{R^3}{A^3}(1 - {\lambda}^3) = \frac{B^3}{A^3} (1 - {{\lambda}_b}^3)  ,\end{equation}
where  $ \lambda \equiv \lambda_2 = r/R$, $ \lambda_a = a/A$ and $\lambda_b = b/B$. 

The Cauchy stress tensor, $\mathbf{T}$,  describes the state of stress in the deformed material. Due to the  symmetry of the problem, there are no shear stresses in the ($\mathbf{e}_i \otimes\mathbf{e}_j$) basis, so that  the only non-zero components of the Cauchy stress tensor $\mathbf{T}$ are $t_1 = T_{11}$ (\textit{radial stress}) and $t_2 = T_{22} = T_{33}$ (\textit{hoop stress}). These are the \textit{principal stresses}. Because the shell is made of an incompressible isotropic hyperelastic material, they are  given by \cite{Ogden}
\begin{equation}\label{stress}
 t_1 =   {\lambda}_1 \frac{\partial W}{\partial {\lambda}_1}- p,   \qquad  t_2 =   {\lambda}_{2} \frac{\partial W}{\partial {\lambda}_{2}}- p,
\end{equation} where $W = W(\lambda_1,{\lambda}_{2},{\lambda}_{3})$ is the strain energy density function and $p$ is a Lagrange multiplier associated with the incompressibility constraint, $\det{\mathbf{ F}} =1 $, to be determined from  the boundary conditions.

In the absence of body forces, the equations for mechanical equilibrium reduce to
\begin{equation}\label{eqmot}\mathrm{div}\ \mathbf{T}=\mathbf{0} .\end{equation}
Computing the divergence in spherical polar coordinates, we find that the only non-trivial component of (\ref{eqmot}) is
\begin{equation}\label{eqmot2} 
\frac{\text dt_1}{\text dr} = \frac{2}{r}(t_2 - t_1).
\end{equation}
Next, we define the auxiliary function $\hat{W}(\lambda) = W({\lambda}^{-2},\lambda,\lambda)$. 
Then we can show that  this equation is equivalent to  \cite{Ogden}
\begin{equation}\label{dt1} 
\frac{\text dt_1}{\text d \lambda} = \frac{ \hat{W}'(\lambda) }{1 - \lambda ^3}  . \end{equation}

We call $P$ the inflation, i.e. the excess of the internal pressure over the external pressure, so that the boundary conditions are $t_1({\lambda}_a)=-P$ and  $t_1({\lambda}_b) = 0$.  
Integrating (\ref{dt1}) and imposing the boundary conditions, we find that
\begin{equation} \label{t1} 
t_1(\lambda)=\int_{{\lambda}_b}^{{\lambda}} \frac{ \hat{W}'(s) }{1 - {s}^3} \mathrm{d}s \quad \mathrm{and} \quad  P=\int_{{\lambda}_a}^{{\lambda}_b} \frac{ \hat{W}'(\lambda) }{1 - {\lambda}^3} \mathrm{d}\lambda,
\end{equation}
where $s$ is a dummy variable. 
Now, introducing the \textit{thickness parameter} $\delta = (B-A)/A$ and noting from (\ref{lambdaa}) that
\begin{equation} 
\lambda_b = 1 - \frac{1 - \lambda_a ^3}{(1+ \delta)^3}, 
\end{equation}
we can expand $P$  in powers of $\delta$ to find
\begin{equation}
P = \delta \frac{ \hat{W}'(\lambda)}{\lambda ^2}+ \frac{\delta ^2}{ 2 \lambda ^4}\left[ \frac{\lambda ^3 - 2}{\lambda} \hat{W}'(\lambda) -  (\lambda ^3 -1)\hat{W}''(\lambda) \right] + O(\delta ^3),
\end{equation}
where $\lambda = \lambda _a [1 + \mathcal O(\delta)]$.
Hence, for thin shells, $P$ can be approximated by
\begin{equation}\label{pexp} 
P = \delta \frac{ \hat{W}'(\lambda)}{\lambda ^2} . 
\end{equation}

Similarly, we can also approximate $P/ t_2$ for thin shells. 
First we note from (\ref{eqmot2}) and (\ref{t1})  that $t_2$ is given by
\begin{equation} 
t_2(\lambda) =  \frac{\lambda}{2} \hat{W}'(\lambda) + \int_{{\lambda}_b}^{{\lambda}} \frac{ \hat{W}'(s) }{1 - {s}^3} \mathrm{d}s.
\end{equation}
Then expanding $P/ t_2$ to first order in $\delta$, leads to the following approximation:
\begin{equation}\label{pt2exp}
\frac{P}{ t_2} = \frac{2}{\lambda  ^3} \delta + \mathcal O(\delta^2) = 2\left(\frac{A}{a}\right)^3 \delta+ \mathcal O(\delta^2),
\end{equation}
showing how the normal stress component is small compared to the circumferential stress components. 
This connection is \emph{universal} and depends only on the geometrical dimensions of the shell.
Clearly, as the shell inflates, the hoop stress will be much greater than the internal pressure. 

Noting that  $t_2$  is equal to the wall tension $T$ divided by the deformed thickness $(B-A)\lambda _1$, we can also recover from (\ref{pt2exp}) the \emph{classical membrane relation}
\begin{equation} \label{tension} T= \frac{Pr}{2} ,
\end{equation}
where $r = a[1 + \mathcal O(\delta)]$. 
This relation has often been used \emph{a priori} to model the inflation of spherical membranes including rubber balloons  \cite{solidmech} and biological soft tissues such as veinous and arterial aneurysms \cite{softtissue}.
As noted by M\"uller and Strehlow \cite{muller}, this is \color{black} directly related to the Young-Laplace law for bubbles of fluid with surface tension. \color{black}


\section{Pressure-stretch curves}
\label{curves}


By choosing a constitutive model through $W$, we can use  the expression (\ref{pexp}) to plot the pressure-stretch curves  for a thin shell subject to inflation. 

In this paper we  compare the performance of two 3-parameter phenomenological models of hyperelasticity. 
The first model, arguably the first such model \emph{ever}, was proposed by Mooney \cite{mooney} in 1940, as
\begin{equation}\label{wn}
W_\text{M} = C_1 ({\lambda}_1 ^n +  {\lambda}_{2}^n +  {\lambda}_{3}^n -3) + C_2({\lambda}_1 ^n {\lambda}_{2}^n +  {\lambda}_{2}^n {\lambda}_{3}^n + {\lambda}_{3}^n {\lambda}_1 ^n -3) ,
\end{equation} where \color{black} $C_1>0,C_2>0,n>0$. \color{black}
Its initial shear modulus can be computed from the general formula \cite{straight}: $\mu = \tilde W''(1)/4$ where $\tilde W(\lambda)=W(\lambda^{-1},\lambda,1)$, as $\mu_\text{M} = (C_1+C_2)n^2/2$.
It covers three popular 2-parameter models for rubber: the Varga \cite{Varga, destrade03} model when $n=1$, the Mooney-Rivlin model when $n=2$ and the one-term Ogden model \cite{Ogden} when $c_2=0$.
The second model is more recent: the so-called Gent-Gent (GG) Model proposed by Pucci and Saccomandi \cite{saccomandi} in 2002.
Its strain energy function reads as  
\begin{equation} \label{GG}
W_\text{GG} =  - c_1 J_m \mathrm{ln}\left(1 - \frac{ \lambda_1 ^2 +  \lambda_2^2 +  \lambda_3^2 - 3}{J_m}\right) + c_2 \mathrm{ln} \left(\frac{\lambda_1 ^2 \lambda_2^2 +  \lambda_2^2 \lambda_3^2 + \lambda_3^2 \lambda_1 ^2}{3} \right),
\end{equation}
where \color{black} $c_1>0,c_2>0,J_m>0$, \color{black} with initial shear modulus $\mu_\text{GG} =2(c_1+c_2/3)$, independent of $J_m$. 
This model relies on the concept of limited chain extensibility and the constant $J_m$ acts as a stiffening parameter: in particular an equi-biaxial deformation such as spherical inflation is limited to the range $0<\lambda<\lambda_m$ where $\lambda_m$ is the real root to the bicubic $\lambda^{-4} + 2\lambda^2=3+J_m$. 
The GG model has been used very successfully to model the behaviour of rubber-like materials \cite{OgSS04}.
It  covers two popular 2-parameter models for rubber, both due to Gent (hence its name): 
the Gent-Thomas model \cite{gent-thomas} when $J_m \to \infty$ and the Gent model \cite{gent-model} when $c_2=0$.

At first we model the pressure-stretch behaviour for an inflated rubber balloon and we leave aside for the time being the data of Osborne  \cite{osborne} for a monkey bladder subject to inflation.
 In the case of rubber, the  pressure-stretch curve quickly reaches a maximum in pressure (a \textit{limit-point instability}) and then it begins to decrease. 
 This behaviour reflects our own experience with inflating toy balloons and is shown by Osborne's data, see Figure \ref{osborne-rubber}.
 However, when we keep inflating a balloon it eventually becomes harder to stretch further and the pressure increases rapidly again, until bursting point.
 Hence  the pressure-stretch curve should have a maximum, followed by a minimum, or in other words, it should have two ascending branches \cite{muller}. 
 In practice, the pressure reaches a maximum, and then the stretch may suddenly `jump' to a larger value on the second ascending branch: this is the so-called \emph{inflation-jump instability} (Osborne's data unfortunately \color{black} stop \color{black} before that point).
 In conclusion, we must limit the range of the parameters in the model to accommodate the constraints of limit-point and inflation-jump instabilities, as dictated by the physics of inflation.

\begin{figure}[hbt]
\centering
\includegraphics[scale=0.3]{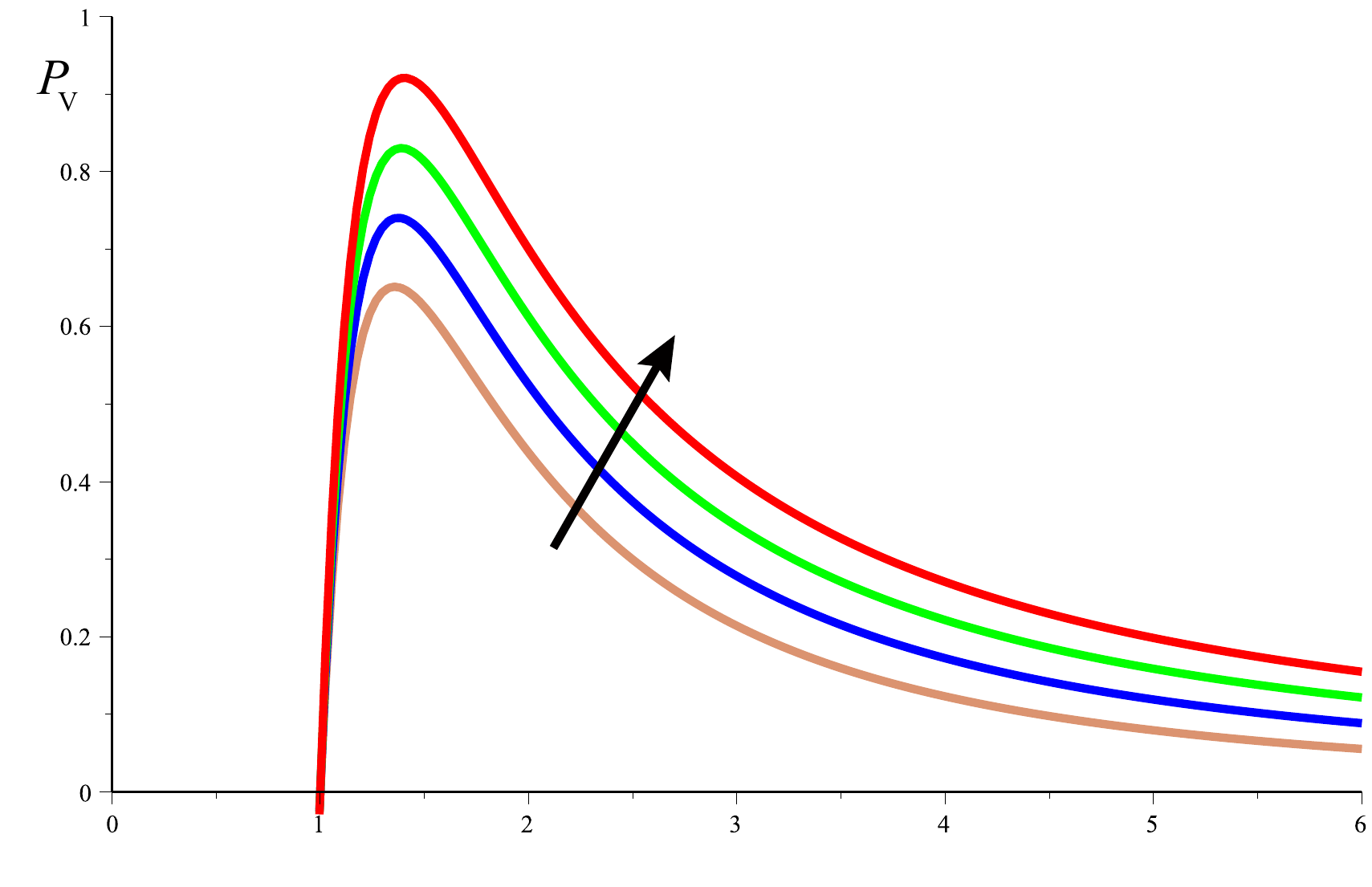}
\includegraphics[scale=0.3]{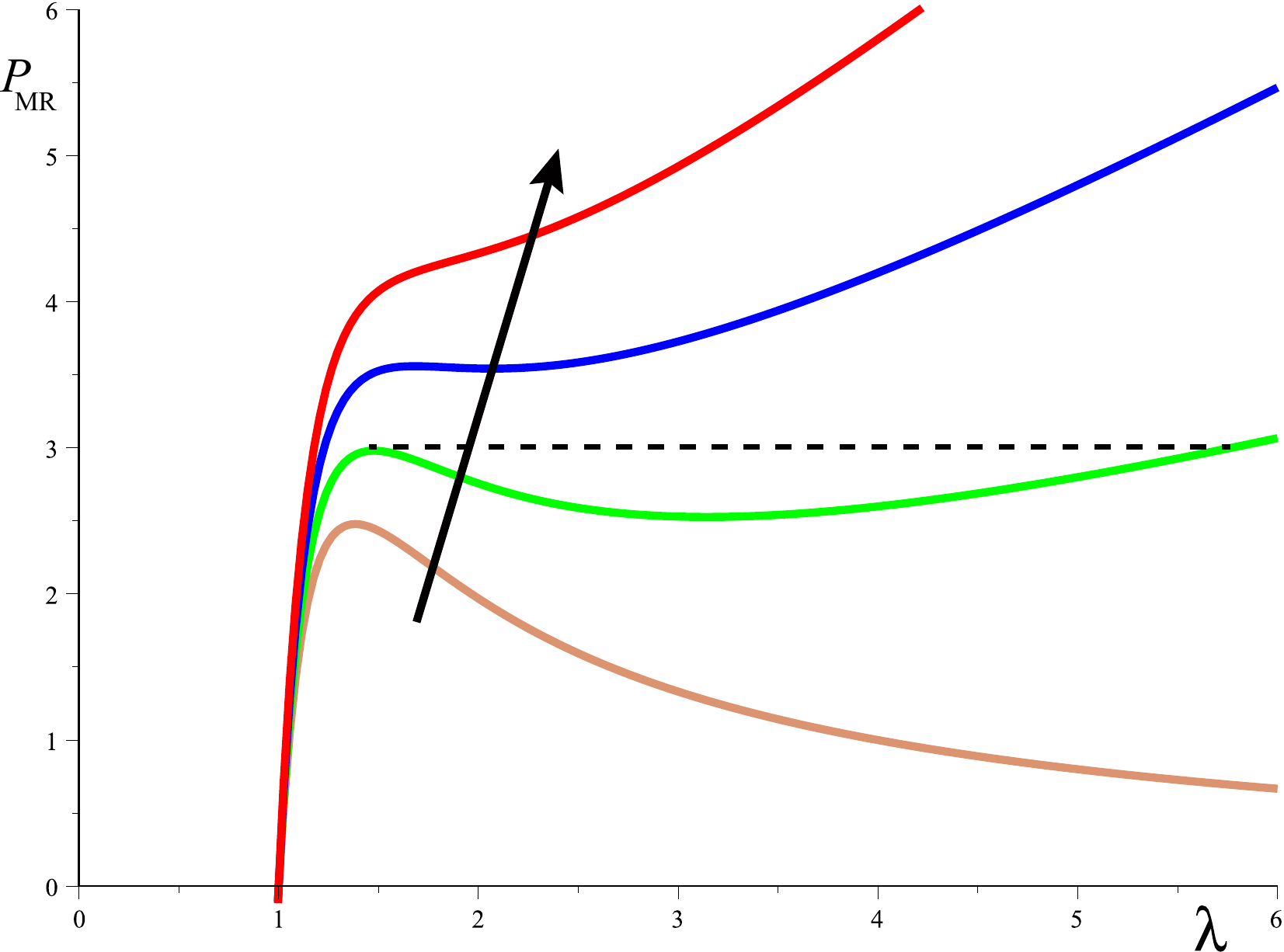}
\includegraphics[scale=0.3]{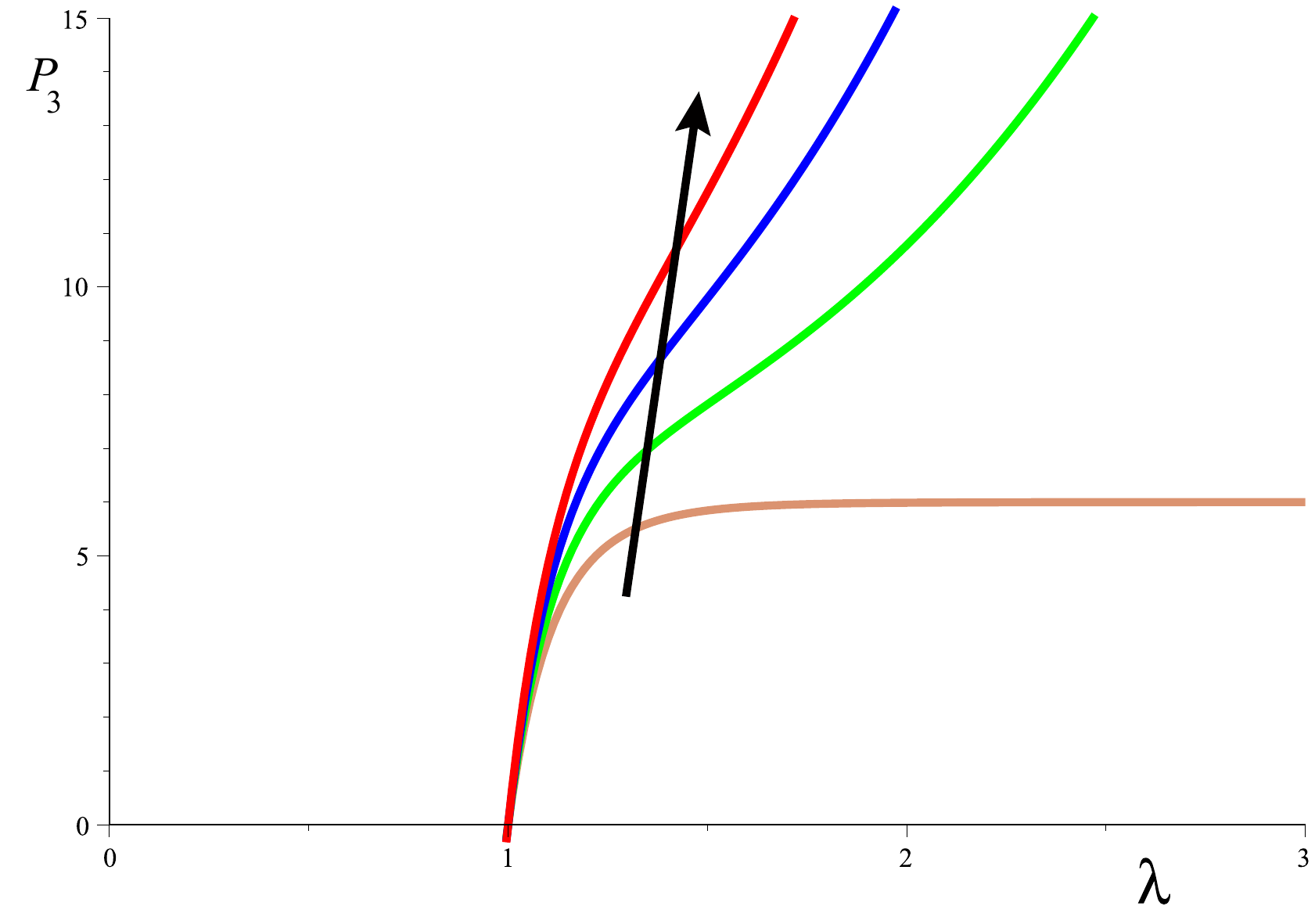}
\caption{Pressure-stretch curve for the inflation of a Mooney balloon. 
For the $n=1$ (Varga), $n=2$ (Mooney-Rivlin) and $n=3$ cases, the material parameter $C_2/C_1$ takes the values $0.0$, $0.1$, $0.2$, $0.3$,  as indicated by the arrows. 
Only the Mooney-Rivlin material (centre) can display both limit-point instability and inflation-point instability (see dashed horizontal line when $C_2/C_1=0.1$).
}
\label{mooneyps}
\end{figure}

Now we examine  the theoretical pressure-stretch curves for the Mooney material, when $W = W_\text{M}$. 
We call $P_\text{M}$ a scaled, non-dimensional measure of the pressure and find, using \eqref{pexp}, that here
\begin{equation}\label{scaledp} 
P_\text{M} \equiv \frac{P}{ C_1 \delta} =  2n \left[\lambda^{n-3} -\lambda^{-2n-3} +  \frac{C_2}{C_1}(\lambda^{2n-3}-\lambda^{-n-3})\right].
\end{equation}
We use this equation to plot several pressure-stretch curves, for the $n=1$ (Varga), $n=2$ (Mooney-Rivlin) and $n=3$ models, and for several values of $C_2/C_1$, see Figure \ref{mooneyps}.

We see that for $n=1$ and $n=3$ the pressure-stretch curves  never exhibit an inflation jump and thus those models must be discarded in the modelling of real rubber balloons.
The Varga model  has a limit-point instability  for all values of $C_2/C_1$; the curve for the model $n=3$ exhibits asymptotic behaviour for $C_2/C_1 = 0$ and is otherwise strictly increasing.
We now explain why out of these three examples only the curve for the $n=2$ Mooney-Rivlin material has two ascending branches. 

To calculate the critical value $(C_2/C_1)_\text{cr}$ of the material parameter $C_2/C_1$ at which the first maximum in the curve ceases to exist, we simply need to determine the stationary point of inflection, by solving the equations 
\begin{equation} \label{crit}
\frac{\text dP}{\text d\lambda} =0, \qquad
\frac{\text d^2P}{\text d\lambda^2} = 0.
\end{equation}
We find two roots for $\lambda_\text{cr}$, the critical value of $\lambda$, but only one of them is real, and is given by
\begin{equation}  
\label{lcr} {\lambda}_\text{cr}=\left[\frac {9 - 7n^2 - 3n\sqrt {5 n^2 - 9}}{(2n-3)(n-3)} \right]^{\frac{1}{3n}}.
\end{equation}
This expression is only well defined when
\begin{equation}
\frac{3}{2} < n < 3.
\end{equation}
Similarly, the corresponding critical value $(C_2/C_1)_\text{cr}$ is only well-defined for $n$ in the same range. 
Its general expression is
\begin{equation} \label{ccr}
(C_2/C_1)_\text{cr}= \frac{(-2n - 3)\left(\frac{-7 n^2 - 3n \sqrt{5n^2-9} +9}{2n^2 - 9n +9}\right)^{-1/3} - (n-3)\left(\frac{-7 n^2 - 3n \sqrt{5n^2-9} +9}{2n^2 - 9n +9}\right)^{2/3}}{(2n-3) \left(\frac{-7 n^2 - 3n \sqrt{5n^2-9} +9}{2n^2 - 9n +9}\right) + n+3}.
\end{equation}

Within that range of possible values for $n$, $P_\text{M} \to \infty$ as $\lambda \to \infty$, ensuring the second ascending branch of the plot.
This range explain why the cases $n=1,3$ must be  disqualified for the modelling of rubber balloons. 
Also, the possibility $(C_2/C_1)_\text{cr}=0$ is excluded from that range, which rules out the one-term Ogden material \cite{destrade}.
As a check, we compute these expressions for the Mooney-Rivlin case $n=2$, and recover the values of Goriely et al. \cite{destrade}: 
\begin{equation} 
 {\lambda}_\text{cr} = (19 + 6 \sqrt{11})^{1/6} \simeq 1.84073, \qquad
 (C_2/C_1)_\text{cr}= \frac{2 \sqrt{11} -3}{5(19+ 6 \sqrt{11})^{1/3}}   \simeq  0.21446.
  \end{equation}
This last quantity explains why the first three curves $C_2/C_1=0.0, 0.1, 0.2$ for $n=2$ in Figure \ref{mooneyps} have a limit-point instability, but the fourth one at $C_2/C_1=0.3$ doesn't.

\begin{figure}[h!]
\centering
\includegraphics[width=0.45\textwidth]{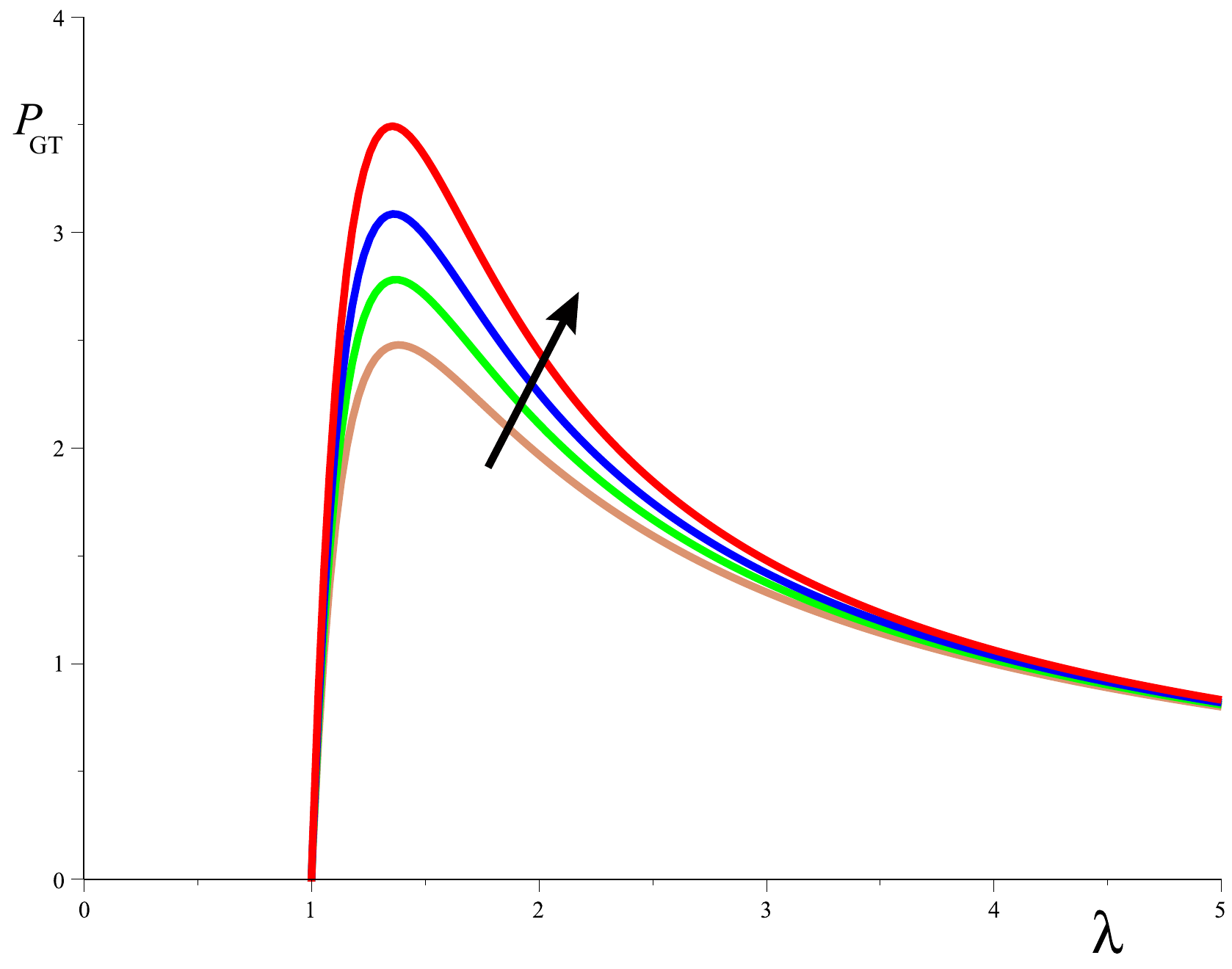}
\includegraphics[width=0.45\textwidth]{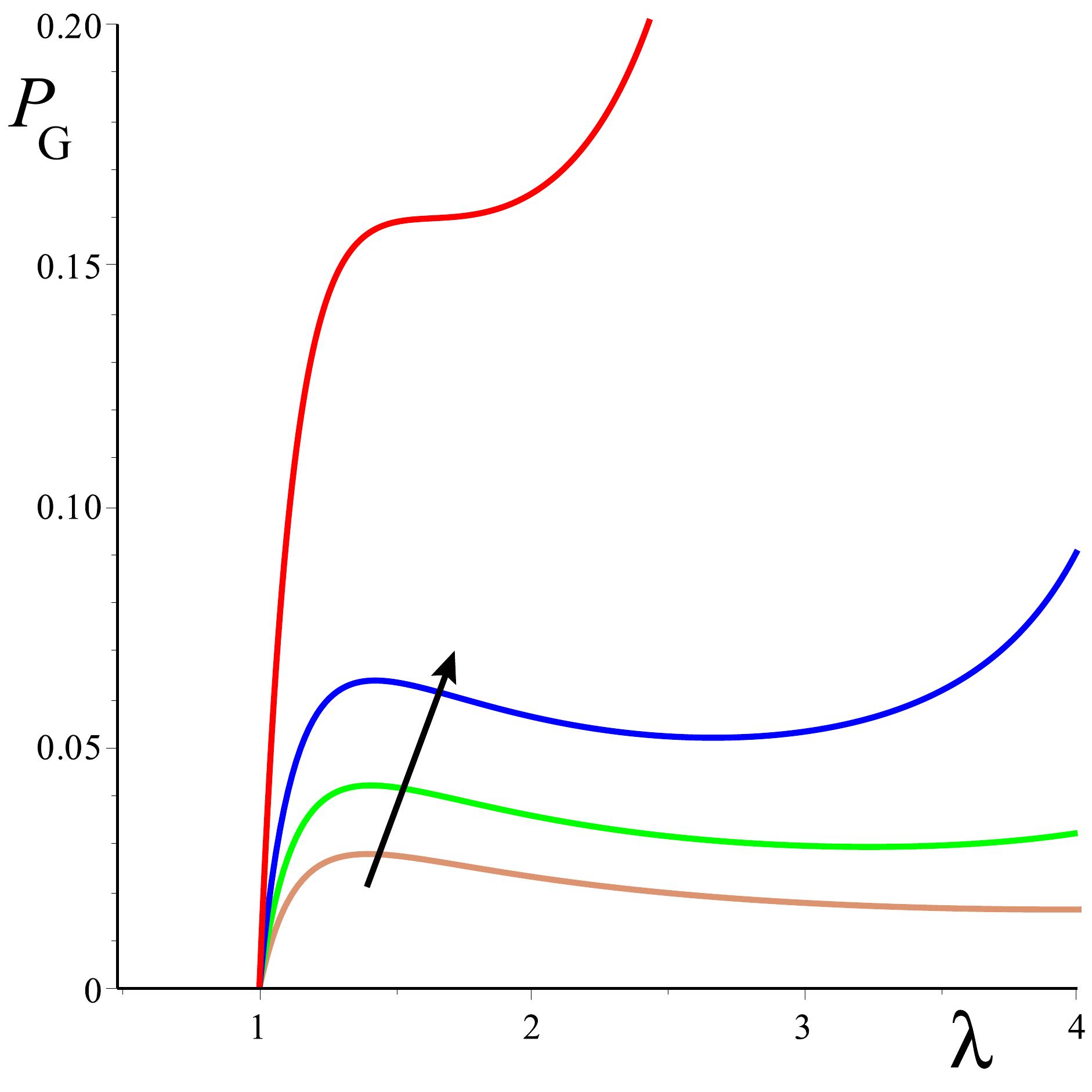}
\caption{Left: Pressure-stretch curves for the inflation of a  Gent-Thomas material,  for the material parameters $c_2/c_1=0.0, 0.3, 0.6, 1.0$ (increasing values indicated by the arrow); Right: Pressure-stretch curves for the inflation of a  Gent material,  for the stiffening parameters $J_m=90.0, 60.0, 40.0, 15.0$ (decreasing values indicated by the arrow).}
\label{gtps}
\end{figure}

Next, we investigate the behaviour of two special cases of the GG model.
We begin with the  Gent-Thomas model \cite{gent-thomas}, obtained when $J_m \to \infty$, as
\begin{equation}
\hat{W}_\text{GT} (\lambda) = c_1 \left(\frac{1}{\lambda ^{4}} + 2  \lambda ^{2} -3 \right) + c_2 \mathrm{  ln}\left( \frac{2}{3 \lambda^2} + \frac{\lambda ^4}{3} \right), 
\end{equation}
and where the corresponding scaled pressure is 
\begin{equation} \label{pgt}  
P_\text{GT} = \frac{P}{c_1 \delta} = 
4(\lambda^{-1} - \lambda^{-7}) + 4\frac{c_2}{c_1}\frac{(\lambda^6 -  1)}{\lambda^3 (\lambda^6 +2)}. 
\end{equation}
We plot the pressure-stretch curves in Figure~\ref{gtps} for various values of $c_2/c_1$. 
We see that it always exhibits a limit-point instability.
These curves illustrate the expected theoretical behaviour, because solving the condition for a local maximum \eqref{crit} gives $\lambda_\text{cr} \simeq 1.182$ and $(c_2/c_1)_\text{cr} = -1.999$.
But this latter value is incompatible with basic physics as it would lead to internal buckling due to loss of ellipticity (it is a simple exercise \cite{destrade04} to show that strong ellipticity for the Gent-Thomas model is equivalent to $c_1>0$, $c_2>0$).
Hence for all physical ratios $c_2/c_1$, the Gent-Thomas model leads to a limit-point instability.
However, its pressure clearly behaves as $\lambda^{-1}$ as $\lambda\to \infty$ so that it can never model a second ascending branch and must thus be discarded as a potential model for rubber balloons.

Finally we consider the Gent model \cite{gent-model}, which is the GG model in the case where $c_2=0$.
We find that the scaled pressure is 
\begin{equation} \label{P-G}
P_\text{G} \equiv \frac{P}{c_1\delta}  = 4\frac{(\lambda^{-1} - \lambda^{-7})J_m}{\left(J_m + 3 - \lambda^{-4} - 2 \lambda^2\right)}, 
\end{equation}
in agreement with Gent \cite{gent99}.
Due to the singularity at $\lambda_m$, there will always be an ascending branch of the curve as $\lambda$ \color{black}increases.
\color{black} 
Solving the equations \eqref{crit}, we find that there is a limit-point instability as long as $J_m > (J_m)_ \text{cr}$, allowing for the modelling of rubber balloons;
In the range $0<J_m<(J_m)_\text{cr}$, the curve has a monotonic increasing behaviour, more appropriate for the modelling of the monkey bladder, see Figure \ref{gtps} for examples.
We find that the critical values of the stiffening parameter and the corresponding critical stretch are \cite{destrade}
\begin{multline}
\lambda_\text{cr} = (10+\sqrt{93})^{1/6} \simeq 1.6426, \\
(J_m)_\text{cr} = \frac{9}{49}(32+\sqrt{93})(10+\sqrt{93})^{1/3} - \simeq 17.638,
\end{multline}
respectively.


\section{Curve fitting}


Here we look at fitting experimental pressure-stretch data for the inflation of a balloon to the theoretical pressure-stretch model (\ref{pexp}) for the Mooney   (\ref{scaledp}) and the GG \eqref{GG} strain energy density functions. 

\begin{figure}[ht!]
                 \includegraphics[width=0.45\textwidth]{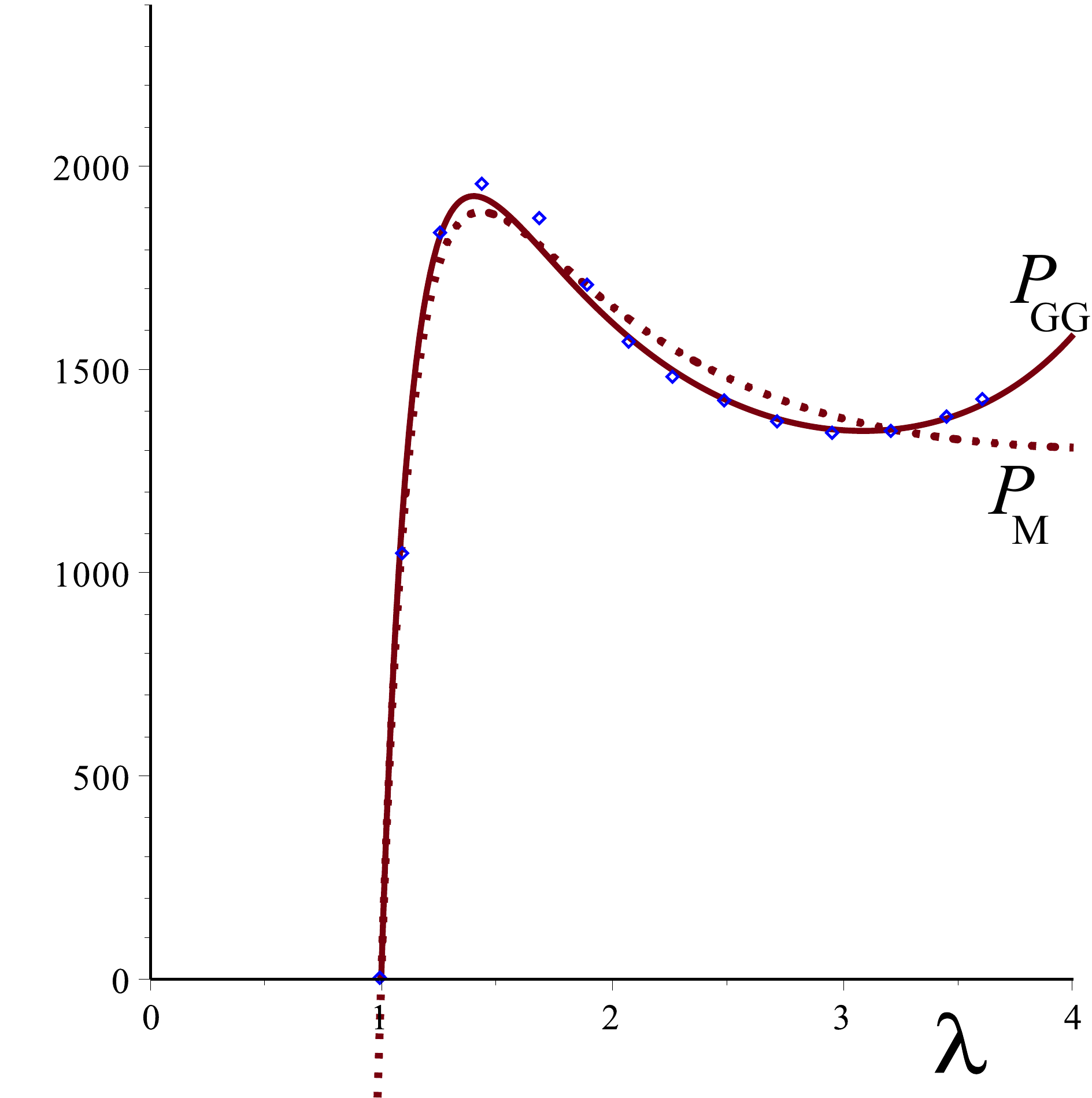}
                \includegraphics[width=0.45\textwidth]{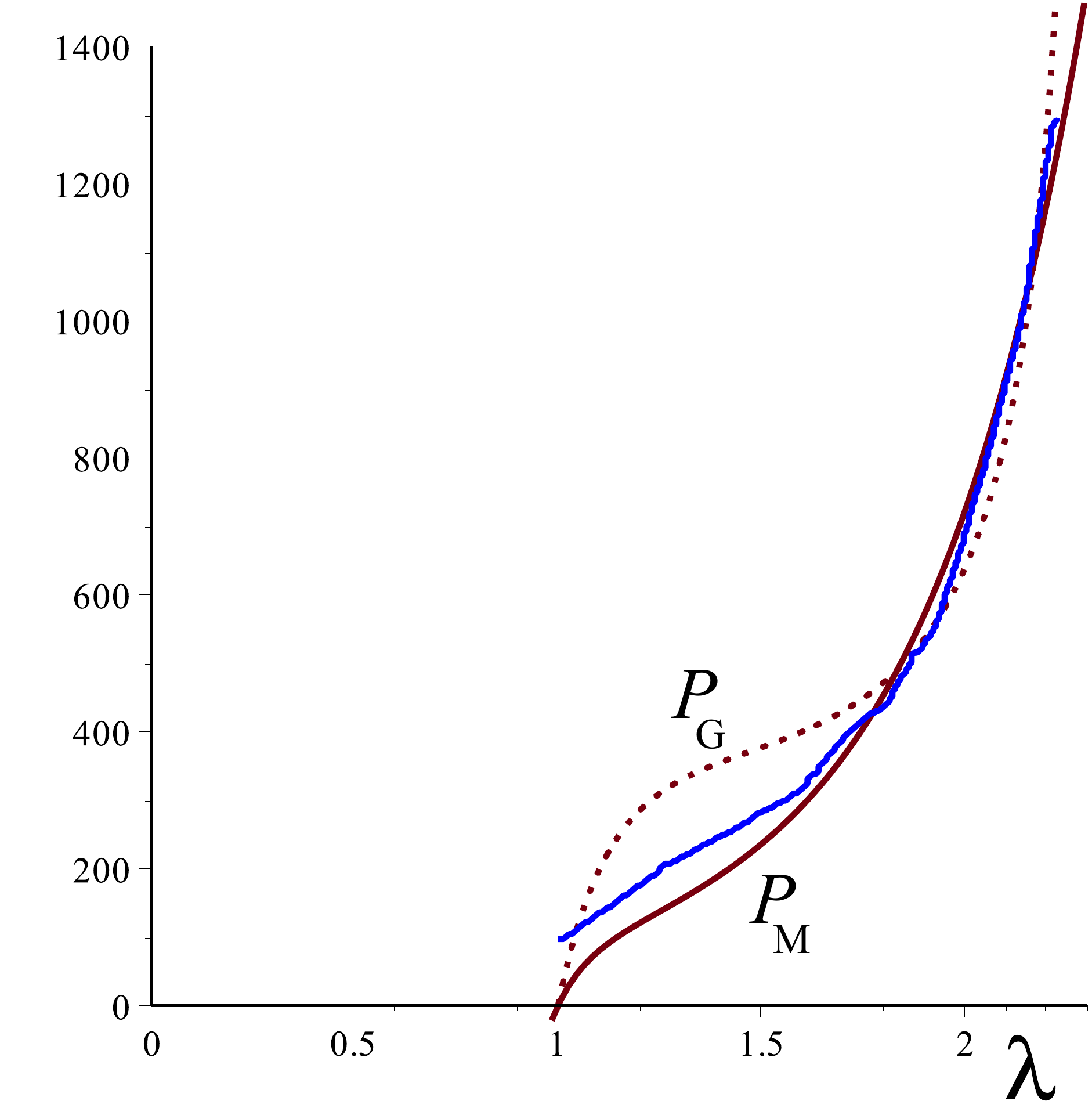}
\caption{Fitted pressure-stretch curves using the Mooney (`$P_\text{M}$') and the Gent-Gent  (`$P_\text{GG}$') models for the inflation of a rubber balloon (left) and a monkey bladder. Blue \color{black}Curve: \color{black} experimental data from \cite{balloon} and \cite{osborne}.}
\label{fitting}
\end{figure}

For the inflation of rubber balloons, we favour the data from Merritt and Weinhaus \cite{balloon} over that of Osborne, because it records the second ascending branch. 
We use their Figure 1, digitised using the \textit{Datathief} \cite{datathief} program. 
It consists of 14 data points. 
The initial radius of the balloon is given as $1.9173$ cm. 
Because the exact thickness of the balloon used  in the experiment is not known, we take it to be that of a typical rubber balloon, 0.5 mm, so that $\delta \simeq 0.0261$. 
Using the non-linear least-squares fitting routine implemented in \emph{Maple} \cite{maple}, we find the values of the best-fit parameters, as listed in Table 1.  
In Figure~\ref{fitting},  we plot the resulting pressure-stretch curves for both models.
\color{black}
In passing we note that the routine used here minimises the absolute error over the whole range of recorded stretches but that another approach, based for instance on minimising the relative error \cite{OgSS04}, could have been adopted. \color{black}

We see that although the two curves behave similarly in the first ascending branch (they have almost the same initial shear modulus: $\mu_\text{M} \simeq 5.5$, $\mu_\text{GG} \simeq 5.8$ kPa), the GG model performs notably better than the Mooney model at moderate to large stretches, and is able to capture perfectly the early rise of the second ascending branch.
Its stiffening parameter $J_m \simeq 53.3$ is compatible with other experiments on rubber \cite{gent-model} and indicates a limiting stretch $\lambda_m \simeq 5.3$. 
The stiffening parameter of the Mooney model is $n \simeq 1.97$, really close to the Mooney-Rivlin model; although it is able to capture the qualitative behaviour of an inflated rubber balloon, it cannot give a good quantitative agreement.

From the point of view of the goodness of fit, using 3-parameter models over 2-parameter models proved to have very little advantage, and we found (not shown here) that the Gent model ($c_2 \equiv 0$) and the Mooney-Rivlin model ($n \equiv 2$) gave almost as good fits as the GG and the Mooney models, respectively. 
Using \color{black} the software package  \emph{R} \cite{r}, \color{black} we verified that the p-values for $c_1$ and $J_m$ are practically zero indicating that these parameters are very significant, while the p-value for $c_2$ is 0.3371 which suggests that this parameter is not significant and that the Gent model (with $c_2 =0$) is sufficient.  
The same conclusion can be reached for the Mooney model as compared to the Mooney-Rivlin model (where $n=2$). 
However, we must recall that from a physical point of view, the dependence of $W$ on the second strain invariant $I_2= \lambda_1^2\lambda_2^2+\lambda_2^2\lambda_3^2+\lambda_3^2\lambda_1^2$ is crucial to model the behaviour of rubber in modes of deformation other than inflation, see Ogden et al.\cite{OgSS04} and Horgan and Smayda \cite{HoSm12}.
The conclusion is that the GG model \color{black} gives the best fitting compared to the other models for the modelling of rubber in spherical inflation. \color{black}

\begin{table}[ht!]
\begin{center}
\begin{tabular}{l*{4}{l}}
Model              &Parameters  &   &    \\
\hline
\hline
Mooney & $C_1 = 26.76$ kPa & $C_2 =1731$ Pa  & $n= 1.968$    \\
Gent-Gent      & $c_1 = 27.65$ kPa & $c_2 =3851$ Pa & $J_m = 53.33$    \\
\hline
Mooney & $C_1 = 2065$ Pa & $C_2 =74.44$ Pa  & $n= 5.03$    \\
Gent-Gent      & $c_1 = 31.32$ kPa & $c_2 =0$ Pa & $J_m = 8.127$    
\end{tabular}
\caption{Best-fit parameters for the fitting of the Mooney and GG models to the pressure-stretch curves for the inflation of a rubber balloon \cite{balloon} (second and third rows) and a monkey bladder \cite{osborne} (fourth and fifth rows).}
\end{center}
\end{table}

Finally, for completeness, we look at fitting Osborne's experimental data \cite{osborne} for the inflation of a monkey bladder.
Unlike the rubber balloon, the pressure-stretch curve has no maximum in the case of the monkey bladder, and exhibits instead a monotonic increasing response.
Osborne's data  was recovered by scanning his Figure 10 and digitising the curve using the \textit{Datathief} \cite{datathief} program. 
The initial radius is given as $1.3347$ cm, and we take the initial thickness to be $2$ mm so that $\delta \simeq 0.15$. 
Here we find that the Mooney model with $n\simeq 5$ gives the best fit, while the GG model is not able to find a best set of parameters when the constraint $c_2 \ge 0$ is enforced. 
It has to settle for the Gent model instead ($c_2=0$) and \color{black} gives \color{black} a poor fitting in the moderate stretch range.
We refer to  Figure~\ref{fitting} and Table 1 for more detail.
However, it must be kept in mind that Osborne's data starts with a non-zero pressure, and that it is thus not possible to locate precisely what should be the initial point at $\lambda=1$.
Also, as noted by Osborne, the bladder tissue exhibits ``complex aeolotropism'' due to the presence of ``a web of elastic fibres with a variable amount of inextensible white fibres intermixed''.  
It follows that this strong anisotropy must be taken into account when modelling the inflation of bladder and that the limit of the assumptions made in this paper have been reached.


\numberwithin{equation}{section}

\appendix


\section{Inflation and extension of a cylinder}
 

Here we establish  the pressure-stretch relationship for a cylindrical shell made of an incompressible istotropic hyperelastic material, subject to internal pressure $P$ and constant axial stretch $\lambda_3$. The derivation is very similar to that of Section \ref{derivation}, see also Ogden \cite{Ogden}. 

 Let \color{black} $\mathbf{X}=\mathbf{X}(R, \Theta, Z)$ and $\mathbf{x}= \mathbf{x}(r, \theta, z)$  \color{black} denote the position of a material particle in the reference and current configurations, respectively. 
 The associated orthonormal bases are $(\mathbf{E}_1,\mathbf{E}_{2},\mathbf{E}_{3})$ and  $(\mathbf{e}_1,\mathbf{e}_{2},\mathbf{e}_{3})$, respectively.  
 Assuming the shell retains its cylindrical symmetry under deformation,  the motion of a particle in the shell can be described by
\begin{equation}\label{motion-cyl}  
r = r(R), \quad \theta = \Theta, \quad z = \lambda_3 Z,
\end{equation}
so that the deformation gradient, $\mathbf{F} = \partial \mathbf{x}/\partial \mathbf{X}$,  is 
\begin{equation}
\mathbf{ F} = \mathrm{diag}(\text dr/\text dR, r/R, \lambda_3), 
\end{equation}
in the ($\mathbf{e}_i \otimes\mathbf{E}_j$) basis. 
Hence, the principal stretches are $\lambda_1 = \text dr/\text dR$ (\textit{radial stretch}), $\lambda_2 = r/R$ (\textit{circumferential stretch}) and $\lambda_3 = \mathrm{constant}$ (\textit{axial stretch}). 
From the incompressibility condition, $\det \mathbf{ F} =1$, we have
\begin{equation}\label{incomp-cyl} 
\frac{\text dr}{\text dR} = \lambda_3^{-1} \frac{R}{r}. 
\end{equation}

Letting $A$, $B$ and $a$, $b$ denote the inner radius and outer radius of the shell in the reference and current configurations, respectively,  and solving (\ref{incomp-cyl}) leads to 
\begin{equation}
\lambda_3 ^{-1} (R^2 - A^2) = r^2 - a^2,
\qquad 
\lambda_3 ^{-1} (R^2 - B^2) = r^2 - b^2, 
\end{equation}
so that
\begin{equation} \label{lambdaa-cyl}
\lambda_3 ^{-1}  - {{\lambda}_a}^2 
= \frac{R^2}{A^2}(\lambda_3 ^{-1} - {\lambda}^2)
 = \frac{B^2}{A^2} (\lambda_3 ^{-1} - {{\lambda}_b}^2), 
\end{equation}
where  $ \lambda \equiv \lambda_2 = r/R$, $ \lambda_a = a/A$ and $\lambda_b = b/B$. 

Due to the  symmetry of the problem,  the only non-zero components of the Cauchy stress tensor $\mathbf{T}$ are $t_1 = T_{11}$(\textit{radial stress}), $t_2 = T_{22}$ (\textit{hoop stress}) and  $t_3 = T_{33}$ (\textit{axial stress}). 
From the equilibrium equation, $\mathrm{div}\ \mathbf{T}=\mathbf{0}$, we have
\begin{equation}\label{eqmot-cyl} 
\frac{\text d t_1}{\text d r} = \frac{1}{r}(t_2 - t_1),
\end{equation}
where $t_1 =   {\lambda}_1 \partial W/\partial {\lambda}_1- p$, $t_2 =   {\lambda}_2 \partial W/\partial {\lambda}_2- p$, and $p$ is a Lagrange multiplier due by the constraint of incompressibility.  
Next, introducing the auxiliary function  $\hat{W}(\lambda, \lambda_3) = W(\lambda^{-1}  \lambda_3^{-1},\lambda,\lambda_3)$ and  using the chain rule, we find
\begin{equation} 
\frac{\partial \hat{W}}{\partial \lambda} = \lambda^{-1} (t_2 - t_1).
\end{equation}
Hence, according to (\ref{eqmot-cyl}), we have
\begin{equation}\label{dt1dr-cyl} 
\frac{\text d t_1}{\text d r} = \frac{\text d \lambda}{\text d r} \frac{\text d t_1}{\text d \lambda} = \frac{\lambda}{r} \frac{\partial \hat{W}}{\partial \lambda}.
\end{equation}
Because $\lambda = r/R$ and $R =\sqrt {\lambda_3 A^2 +\lambda_3  r^2 - a^2}$, we find that
\begin{equation}\label{dlambda-cyl} 
\frac{\text d \lambda}{\text d r}= \frac{1}{R}(1- \lambda ^2 \lambda_3).
\end{equation}
Hence, from  (\ref{dt1dr-cyl}) and (\ref{dlambda-cyl}), we deduce the  expression
\begin{equation}
\label{dt1-cyl} 
\frac{\text d t_1}{\text d \lambda} = \frac{\partial \hat{W}}{\partial \lambda}  \frac{1}{1 - \lambda ^2 \lambda_3}.
\end{equation}
Integrating (\ref{dt1-cyl}) and imposing the boundary conditions $t_1({\lambda}_a)=-P$ and  $t_1({\lambda}_b) = 0$, we find that
\begin{equation} \label{t1-cyl} 
t_1(\lambda)=\int_{{\lambda}_b}^{{\lambda}}\frac{1}{1 - {s}^2 \lambda_3} \frac{ \partial \hat{W}}{\partial s}  \mathrm{d}s 
\quad \mathrm{and} \quad  
P = \int_{{\lambda}_a}^{{\lambda_b}}\frac{1}{1 - {\lambda}^2 \lambda_3} \frac{ \partial \hat{W}}{\partial \lambda}  \mathrm{d}\lambda ,\end{equation}
where $s$ is a dummy variable.
Expanding $P$ in the thickness parameter $\delta = (B-A)/A$ leads to
\begin{equation}P = \frac{1}{ \lambda \lambda_3} \frac{ \partial \hat{W}}{\partial \lambda} \delta - \frac{1}{2 \lambda ^3 \lambda_3 ^2} \left[\lambda ^3 \lambda_3  \frac{ \partial ^2 \hat{W}}{\partial \lambda ^2} - \lambda \frac{ \partial ^2 \hat{W}}{\partial \lambda ^2} +  \frac{ \partial \hat{W}}{\partial \lambda} \right] \delta^2 + \mathcal O(\delta ^3) ,
\end{equation}
where $\lambda = \lambda _a [1 + \mathcal O(\delta)]$. 

Next we set $\lambda_3 = 1$, so that the cylinder is  deforming in the radial direction only. 
Determining $t_2$ using (\ref{eqmot-cyl}) and (\ref{t1-cyl}), and  expanding $P/t_2$ to first order in $\delta$ leads to the following approximation for thin shells:
\begin{equation} \label{pt2-cyl} \frac{P}{t_2} = \frac{1}{ \lambda ^2} \delta . \end{equation}
Because $t_2$ is equal to the wall tension $T$ divided by the deformed  $(B-A)\lambda _1$, we  can recover from (\ref{pt2-cyl}) the classical membrane  relation
\begin{equation} T = Pr, \end{equation}
where $r = a[1+ \mathcal O(\delta)]$. 
This is the pendant formula to  (\ref{tension}) for cylinders \cite{softtissue}. 


\end{document}